

%

%
 \font\twelvebf=cmbx12
 \font\twelvett=cmtt12
 \font\twelveit=cmti12
 \font\twelvesl=cmsl12
 \font\twelverm=cmr12		\font\ninerm=cmr9
 \font\twelvei=cmmi12		\font\ninei=cmmi9
 \font\twelvesy=cmsy10 at 12pt	\font\ninesy=cmsy9
 \skewchar\twelvei='177		\skewchar\ninei='177
 \skewchar\seveni='177	 	\skewchar\fivei='177
 \skewchar\twelvesy='60		\skewchar\ninesy='60
 \skewchar\sevensy='60		\skewchar\fivesy='60
%
%

%
 \font\fourteenrm=cmr12 scaled 1200
 \font\seventeenrm=cmr12 scaled 1440
 \font\fourteenbf=cmbx12 scaled 1200
 \font\seventeenbf=cmbx12 scaled 1440
%
%

%
%
%
\font\tenmsb=msbm10
\font\twelvemsb=msbm10 scaled 1200
\newfam\msbfam

%
\font\tensc=cmcsc10
\font\twelvesc=cmcsc10 scaled 1200
\newfam\scfam

%
\def\seventeenpt{\def\rm{\fam0\seventeenrm}%
 \textfont\bffam=\seventeenbf	\def\bf{\fam\bffam\seventeenbf}}
\def\fourteenpt{\def\rm{\fam0\fourteenrm}%
 \textfont\bffam=\fourteenbf	\def\bf{\fam\bffam\fourteenbf}}
\def\twelvept{\def\rm{\fam0\twelverm}%
 \textfont0=\twelverm	\scriptfont0=\ninerm	\scriptscriptfont0=\sevenrm
 \textfont1=\twelvei	\scriptfont1=\ninei	\scriptscriptfont1=\seveni
 \textfont2=\twelvesy	\scriptfont2=\ninesy	\scriptscriptfont2=\sevensy
 \textfont3=\tenex	\scriptfont3=\tenex	\scriptscriptfont3=\tenex
 \textfont\itfam=\twelveit	\def\it{\fam\itfam\twelveit}%
 \textfont\slfam=\twelvesl	\def\sl{\fam\slfam\twelvesl}%
 \textfont\ttfam=\twelvett	\def\tt{\fam\ttfam\twelvett}%
 \scriptfont\bffam=\tenbf 	\scriptscriptfont\bffam=\sevenbf
 \textfont\bffam=\twelvebf	\def\bf{\fam\bffam\twelvebf}%
 \textfont\scfam=\twelvesc	\def\sc{\fam\scfam\twelvesc}%
 \textfont\msbfam=\twelvemsb	
 \baselineskip 14pt%
 \abovedisplayskip 7pt plus 3pt minus 1pt%
 \belowdisplayskip 7pt plus 3pt minus 1pt%
 \abovedisplayshortskip 0pt plus 3pt%
 \belowdisplayshortskip 4pt plus 3pt minus 1pt%
 \parskip 3pt plus 1.5pt
 \setbox\strutbox=\hbox{\vrule height 10pt depth 4pt width 0pt}}
\def\tenpt{\def\rm{\fam0\tenrm}%
 \textfont0=\tenrm	\scriptfont0=\sevenrm	\scriptscriptfont0=\fiverm
 \textfont1=\teni	\scriptfont1=\seveni	\scriptscriptfont1=\fivei
 \textfont2=\tensy	\scriptfont2=\sevensy	\scriptscriptfont2=\fivesy
 \textfont3=\tenex	\scriptfont3=\tenex	\scriptscriptfont3=\tenex
 \textfont\itfam=\tenit		\def\it{\fam\itfam\tenit}%
 \textfont\slfam=\tensl		\def\sl{\fam\slfam\tensl}%
 \textfont\ttfam=\tentt		\def\tt{\fam\ttfam\tentt}%
 \scriptfont\bffam=\sevenbf 	\scriptscriptfont\bffam=\fivebf
 \textfont\bffam=\tenbf		\def\bf{\fam\bffam\tenbf}%
 \textfont\scfam=\tensc		\def\sc{\fam\scfam\tensc}%
 \textfont\msbfam=\tenmsb	
 \baselineskip 12pt%
 \abovedisplayskip 6pt plus 3pt minus 1pt%
 \belowdisplayskip 6pt plus 3pt minus 1pt%
 \abovedisplayshortskip 0pt plus 3pt%
 \belowdisplayshortskip 4pt plus 3pt minus 1pt%
 \parskip 2pt plus 1pt
 \setbox\strutbox=\hbox{\vrule height 8.5pt depth 3.5pt width 0pt}}

%
\def\twelvepoint{%
 \def\small{\tenpt\rm}%
 \def\normal{\twelvept\rm}%
 \def\big{\fourteenpt\rm}%
 \def\huge{\seventeenpt\rm}%
 \footline{\hss\twelverm\folio\hss}
 \normal}
%

%
\def\bigbold{\big\bf}

%
\catcode`\@=11
%
%
\def\footnote#1{\edef\@sf{\spacefactor\the\spacefactor}#1\@sf
 \insert\footins\bgroup\small
 \interlinepenalty100	\let\par=\endgraf
 \leftskip=0pt		\rightskip=0pt
 \splittopskip=10pt plus 1pt minus 1pt	\floatingpenalty=20000
 \smallskip\item{#1}\bgroup\strut\aftergroup\@foot\let\next}
%
%
%
%
\def\hexnumber@#1{\ifcase#1 0\or 1\or 2\or 3\or 4\or 5\or 6\or 7\or 8\or
 9\or A\or B\or C\or D\or E\or F\fi}
\edef\msbfam@{\hexnumber@\msbfam}

%
%
%
\catcode`\@=12

\newcount\EQNO      \EQNO=0
\newcount\FIGNO     \FIGNO=0
\newcount\REFNO     \REFNO=0
\newcount\SECNO     \SECNO=0
\newcount\SUBSECNO  \SUBSECNO=0
\newcount\FOOTNO    \FOOTNO=0
\newbox\FIGBOX      \setbox\FIGBOX=\vbox{}
\newbox\REFBOX      \setbox\REFBOX=\vbox{}
\newbox\RefBoxOne   \setbox\RefBoxOne=\vbox{}

\expandafter\ifx\csname normal\endcsname\relax\def\normal{\null}\fi

\def\Eqno{\global\advance\EQNO by 1 \eqno(\the\EQNO)%
    \gdef\label##1{\xdef##1{\nobreak(\the\EQNO)}}}
\def\Fig#1{\global\advance\FIGNO by 1 Figure~\the\FIGNO%
    \global\setbox\FIGBOX=\vbox{\unvcopy\FIGBOX
      \narrower\smallskip\item{\bf Figure \the\FIGNO~~}#1}}
\def\Ref#1{\global\advance\REFNO by 1 \nobreak[\the\REFNO]%
    \global\setbox\REFBOX=\vbox{\unvcopy\REFBOX\normal
      \smallskip\item{\the\REFNO .~}#1}%
    \gdef\label##1{\xdef##1{\nobreak[\the\REFNO]}}}
\def\Section#1{\SUBSECNO=0\advance\SECNO by 1
    \bigskip\leftline{\bf \the\SECNO .\ #1}\nobreak}
\def\Subsection#1{\advance\SUBSECNO by 1
    \medskip\leftline{\bf \ifcase\SUBSECNO\or
    a\or b\or c\or d\or e\or f\or g\or h\or i\or j\or k\or l\or m\or n\fi
    )\ #1}\nobreak}
\def\Footnote#1{\global\advance\FOOTNO by 1
    \footnote{\nobreak$\>\!{}^{\the\FOOTNO}\>\!$}{#1}}
\def\SameFootnote{$\>\!{}^{\the\FOOTNO}\>\!$}

\def\References{\bigskip\centerline{\bf REFERENCES}
                \smallskip\copy\REFBOX}
\def\NewRefPage{\setbox\RefBoxOne=\vbox{\unvcopy\REFBOX}
		\setbox\REFBOX=\vbox{}
		\def\References{\bigskip\centerline{\bf REFERENCES}
                		\nobreak\smallskip\nobreak\copy\RefBoxOne
				\vfill\eject
				\smallskip\copy\REFBOX}
		\def\NewRefPage{}}




\font\tenbm=cmmib10
\font\ninei=cmmi9
\newfam\bmfam

\def\tenpointbmit{
\textfont\bmfam=\tenbm
\scriptfont\bmfam=\seveni
\scriptscriptfont\bmfam=\fivei
\def\bmit{\fam\bmfam\tenbm}
}

\tenpointbmit

\mathchardef\Gamma="7100
\mathchardef\Delta="7101
\mathchardef\Theta="7102
\mathchardef\Lambda="7103
\mathchardef\Xi="7104
\mathchardef\Pi="7105
\mathchardef\Sigma="7106
\mathchardef\Upsilon="7107
\mathchardef\Phi="7108
\mathchardef\Psi="7109
\mathchardef\Omega="710A
\mathchardef\alpha="710B
\mathchardef\beta="710C
\mathchardef\gamma="710D
\mathchardef\delta="710E
\mathchardef\epsilon="710F
\mathchardef\zeta="7110
\mathchardef\eta="7111
\mathchardef\theta="7112
\mathchardef\iota="7113
\mathchardef\kappa="7114
\mathchardef\lambda="7115
\mathchardef\mu="7116
\mathchardef\nu="7117
\mathchardef\xi="7118
\mathchardef\pi="7119
\mathchardef\rho="711A
\mathchardef\sigma="711B
\mathchardef\tau="711C
\mathchardef\upsilon="711D
\mathchardef\phi="711E
\mathchardef\cho="711F
\mathchardef\psi="7120
\mathchardef\omega="7121
\mathchardef\varepsilon="7122
\mathchardef\vartheta="7123
\mathchardef\varpi="7124
\mathchardef\varrho="7125
\mathchardef\varsigma="7126
\mathchardef\varphi="7127



%
%
\twelvepoint			
%
%



\font\tenbm=cmmib10
\font\ninei=cmmi9
\newfam\bmfam

\def\tenpointbmit{
\textfont\bmfam=\tenbm
\scriptfont\bmfam=\seveni
\scriptscriptfont\bmfam=\fivei
\def\bmit{\fam\bmfam\tenbm}
}

\tenpointbmit

\mathchardef\Gamma="7100
\mathchardef\Delta="7101
\mathchardef\Theta="7102
\mathchardef\Lambda="7103
\mathchardef\Xi="7104
\mathchardef\Pi="7105
\mathchardef\Sigma="7106
\mathchardef\Upsilon="7107
\mathchardef\Phi="7108
\mathchardef\Psi="7109
\mathchardef\Omega="710A
\mathchardef\alpha="710B
\mathchardef\beta="710C
\mathchardef\gamma="710D
\mathchardef\delta="710E
\mathchardef\epsilon="710F
\mathchardef\zeta="7110
\mathchardef\eta="7111
\mathchardef\theta="7112
\mathchardef\iota="7113
\mathchardef\kappa="7114
\mathchardef\lambda="7115
\mathchardef\mu="7116
\mathchardef\nu="7117
\mathchardef\xi="7118
\mathchardef\pi="7119
\mathchardef\rho="711A
\mathchardef\sigma="711B
\mathchardef\tau="711C
\mathchardef\upsilon="711D
\mathchardef\phi="711E
\mathchardef\cho="711F
\mathchardef\psi="7120
\mathchardef\omega="7121
\mathchardef\varepsilon="7122
\mathchardef\vartheta="7123
\mathchardef\varpi="7124
\mathchardef\varrho="7125
\mathchardef\varsigma="7126
\mathchardef\varphi="7127



\centerline{\bigbold AN EINSTEIN-HILBERT ACTION}\vskip 0.7cm
\centerline{\bigbold  FOR }\vskip 0.7cm
\centerline{\bigbold AXI-DILATON GRAVITY IN 4-DIMENSIONS}
\bigskip\bigskip\bigskip

\centerline{T Dereli${}^{\dagger}$}
\medskip
\centerline{Robin W Tucker}
\medskip

\centerline{\it School of Physics and Materials,}
\centerline{\it University of Lancaster,
		Bailrigg, Lancs. LA1 4YB, UK}
\centerline{\tt rwt{\rm @}lavu.physics.lancaster.ac.uk}

\vskip 1cm
\vskip 2cm

\bigskip\bigskip\bigskip\bigskip

\centerline{\bf ABSTRACT}
\vskip 1cm

\midinsert
\narrower\narrower\noindent


We examine the axi-dilatonic sector of low energy
string theory and demonstrate how the gravitational interactions involving
the axion and dilaton fields may be derived from a geometrical action principle
involving the curvature scalar associated with a non-Riemannian connection.
In this geometry the antisymmetric tensor 3-form field  determines the
torsion of the connection on the frame bundle while the gradient of the
metric is determined by the dilaton field.
By expressing the theory in terms of the Levi-Civita connection associated
with the metric in the ``Einstein frame'' we
 confirm that the field equations derived from the non-Riemannian
Einstein-Hilbert action  coincide with the axi-dilaton sector of the
low energy effective action derived from string theory.

\endinsert




\vfill

${}^{\dagger}$
{ Mathematics Department, Middle East Technical University, Ankara, Turkey}

\eject
\headline={\hss\rm -~\folio~- \hss}     

\def\frac#1#2{{#1\over #2}}

\Section{Introduction}

\def\wd{\wedge}


\def\ax{{\cal A}}
\def\R#1#2{R^#1{}_#2}
\def\q#1#2{q^#1{}_#2}
\def\Q#1#2{Q^#1{}_#2}
\def\Om#1#2{\Omega^#1{}_#2}
\def\om#1#2{\omega^#1{}_#2}
\def\L#1#2{\Lambda^#1{}_#2}
\def\K#1#2{K^#1{}_#2}
\def\et#1#2{\eta^#1{}_#2}
\def\e#1{e^#1{}}
\def\sp#1{#1^{\prime}}
\def\l{\lambda}
\def\gg{{\cal G}}

In the absence of matter
Einstein's theory of gravity can be derived from a
geometrical action principle. The action density may be expressed elegantly
in terms of the curvature scalar associated with the  curvature of the
Levi-Civita connection.
Such a connection $\nabla$
 is torsion-free and metric compatible.
Thus for all vector fields $X,Y$ on the spacetime manifold:
${\bf T}(X,Y)\equiv\nabla _X Y-\nabla _Y X -[X,Y]=0 $
and
${\bf Q}\equiv\nabla g =0$ where
 $ g$ denotes the metric tensor, $\bf T$ the torsion tensor of $\nabla$
and ${\bf Q}$ the gradient tensor of $g$ with respect to $\nabla$.
The Levi-Civita connection provides a useful reference
connection since it depends entirely on the metric structure of the
manifold.
Such a metric structure can be used to construct additional terms in the
action describing the interaction of matter with gravity although one needs
a guiding principle in order to remove the adhoc nature of this construction.
Such interactions
can sometimes be derived from more general connections. Torsional
connections have been shown to accommodate gravitational interactions between
spinors \Ref{D Deser, B Zumino Phys. Letts. {\bf B62} (1976) 335}
 as well as scalar fields \Ref{T Dereli, R W Tucker Phys.
Letts {\bf 110B} (1982) 206}\label{\brans }.
 In these approaches the theory can be rewritten in terms of the
Levi-Civita connection so that the torsion and non-metricity tensors can
then be interpreted as matter induced couplings for Einsteinian gravity,
\Ref{ F W Hehl,E Lord, L L Smalley, Gen. Rel. Grav. {\bf 13} (1981) 1037},
\Ref{P Baekler, F W Hehl, E W Mielke ``Non-Metricity and Torsion'' in Proc.
of 4th Marcel Grossman Meeting on General Relativity, Part A, Ed. R Ruffini
(North Holland 1986) 277}, \Ref{V N Ponomariev, Y Obukhov, Gen. Rel. Grav.
{\bf 14} (1982) 309}, \Ref{ J D McCrea, Clas. Q. Grav. {\bf 9} (1992) 553},
\Ref{ A A Coley, Phys. Rev. {\bf D27} (1983) 728}, \Ref{ A A Coley, Phys.
Rev. {\bf D28} (1983) 1829, 1844}, \Ref{ A A Coley, Nuov. Cim. {\bf 69B}
(1982) 89}, \Ref{M Gasperini, Class. Quant. Grav. {\bf 5} (1988) 521},
\Ref{J Stelmach, Class. Quant. Grav. {\bf 8} (1991) 897}, \Ref{ A K
Aringazin, A L Mikhailov, Class. Q. Grav. {\bf 8} (1991) 1685}.  An
alternative approach is to induce matter couplings from string theory.  Low
energy string theory predicts specific couplings between fields for
massless excitations in space-time.  Besides the graviton, gauge fields and
hypothetical gravitinos one must consider the antisymmetric tensor field
and dilaton.  Depending on the topology of the compactification of internal
dimensions there appear further scalar excitations with well defined
coupling schemes.  The pattern of all these couplings gives rise to new
kinds
 of ``duality'' symmetries
that may be used to discover
new kinds of dilatonic black holes in 4-dimensions.
These symmetries may be expressed either in terms of the geometry  of the
metric of the string graviton  or in terms of a Weyl scaled
metric (the so called ``Einstein frame'').

In this note we concentrate on the axi-dilatonic sector of low energy
string theory and demonstrate how the gravitational interactions involving
the axion and dilaton may be derived from a geometrical action principle
involving the curvature scalar associated with a non-Riemannian connection.

\Section{ The Curvature Scalar}

The traditional metric
variation of the Einstein-Hilbert action assumes that the
connection is torsion free and metric compatible
\Ref{A Einstein, S. B.
preuss. Akad. Wiss., 414, (1925)}.
A simpler variational
approach  relaxes the torsion free condition but maintains
the constraint of metric compatibility
\Ref{ T Dereli, R W Tucker, Class. Q. Grav. {\bf 4} (1987) 791}.
In the absence of matter varying
 such a connection and metric (by varying a set of orthonormal coframe fields)
yields the same result.
Alternatively varying a metric and a connection that is constrained to be
torsion free but not metric compatible
yields  the same (pseudo-) Riemannian geometry \Ref{J L Anderson,
{\bf Principles of Relativity Physics}, Academic Press, (1967)
}.
The results of such traditional constrained variations can be unified using
an action principle
in which the structure equations for the torsion and non-metricity
 appear as constraints \Ref{T Dereli, R W Tucker,
 Class. Q. Grav. {\bf 11} (1994) 2575}. This approach has the virtue that it
can accommodate consistent variations of the metric, connection, torsion and
non-metricity forms.

In the following it will prove convenient to choose a g-orthonormal coframe
in which $g$ takes the
form $g=\eta_{ab} e^a\otimes e^b$ with $\eta_{ab}=\pm 1$.
In this coframe we define the connection 1-forms $\L ab$ by
$$\nabla_{X_a}e^c=-\Lambda^c{}_b(X_a)\, e^b\Eqno$$
It follows that the torsion and curvature forms of $\nabla$ are given by
$$T^a=De^a\equiv de^a +\Lambda^a{}_b\wd e^b\Eqno$$
$$R^a{}_b (\Lambda)=
d\Lambda^a{}_b+\Lambda^a{}_c\wd \Lambda^c{}_b\Eqno$$\label{\curv }
We shall often use the exterior covariant derivative $D$ whose general
definition can be found in
\Ref{ I M Benn, R W Tucker, {\bf An Introduction to Spinors and Geometry with
Applications in Physics}, (Adam Hilger) (1987)}\label\book .
We induce a variation of the curvature scalar
${\cal R}$
by varying consistently  the set of g-orthonormal coframe fields $\{e^a\}$
and
the set of connection 1-forms $\L ab$ representing $\nabla$ in such a coframe.
The curvature scalar
${\cal R}$ is defined as the coefficient of the canonical volume form $*1$
in the 4-form
$\R ab (\Lambda ) \wd *(e_a\wd \e b)$ where $*$ denotes the Hodge map.
The torsion tensor can be written in terms of a set of 2-forms $\{T^a\}$
with respect to the coframe $\{e^a\}$:
$$T^a(X,Y)=\frac12 e^a({\bf T}(X,Y))\Eqno$$
The symmetric non-metricity forms $Q_{ab}$ are tensorial and
 $$D\eta_{ab}=-2 Q_{ab}$$

We shall work in the so called ``Einstein frame'' and denote the dilaton field
by $\phi$, (with $q=e^{-\phi}$) and the
axion field by $\ax$.
$\ax$ is related to the
antisymmetric tensor 3-form field  $H$ by the relation:
$$q^2 *H=d{\cal A}$$\label\axion
In terms of these fields we assume that the connection $\nabla$ has torsion
$$T_a=\gg (q) i_a H\Eqno$$\label\tor
and non-metricity
$$\Q ab=\et ab d(f(q))\Eqno$$\label\nonmet
in terms of real functions $f$ and $\gg$.
We may decompose the connection forms
$$\L ab=\Om ab +\q ab +\Q ab\Eqno$$\label\conn
where
$$\q ab=-(i_a Q_{bc}) \e c +(i_b Q_{ac}) \e c\eqno$$
$$\Om ab=\om ab+\K ab\Eqno$$
The Levi-Civita connection forms $\om ab$ are  defined by
$d e^a +\om ab\wd e^b =0$
and
$K_{ab}=\frac{\gg }{2} i_ai_b H$
are the contorsion
forms defined by
$T^a=K^a{}_b\wd e^b$.
Corresponding to the decomposition \conn\  of the connection
one finds
$$\R ab(\Lambda)=\R ab(\Omega) +D(\Omega)(\q ab+\Q ab)+\q ac\wd\q cb +\Q
ac\wd\Q cb
+\q ac\wd\Q cb +\Q ac\wd \q cb$$
Evaluating the exterior covariant derivative and using \tor\  and \nonmet\
one finds after discarding exact forms:
$$\R ab(\Lambda )\wd *(e_a\wd \e b)=\R ab(\omega)\wd *(e_a\wd \e b) -\frac 32
\gg ^2(q) H\wd *H -6 df(q) \wd * df(q) $$\label\action
With the choice
$ \gg (q)= q$, $ f(q)=\sqrt(1/6) ln(q)$
 one recognises   precisely the action for axi-dilaton gravity in the
``Einstein frame''.
This observation suggests that this sector of the theory can be obtained
from a variational principle in which the Lagrangian density is the
curvature scalar of the above connection. Such a variational procedure must
yield the correct field equations for the axion and dilaton fields when
the theory is
re-expressed in terms of the standard (pseudo-)Riemannian geometry.

\Section{ The Einstein-Hilbert Variational Problem for a non-Riemannian
Geometry}

To consistently vary the curvature scalar when the torsion and
non-metricity are constrained by \tor\ and \nonmet\ introduce
multiplier 2-forms $\lambda_a$ and 3-forms $\rho^a{}_b$. We may use these
to impose the torsion and non-metricity constraints in the variational
procedure. Since we are disregarding
Maxwell and Chern-Simons terms in the effective action we shall also assume
that $H$ is a closed form. This is accommodated by introducing a further
0-form multiplier $\mu$ and writing the action as:

$${\cal L}[\Lambda,e,\lambda,\rho,\mu,q,H]=
{\cal R}*1 +\lambda_a \wd (d\e a +\L ab\wd \e b
-\gg (q) i^a H) + \rho^a{}_b\wd (\Q ba -\et ba df(q)) +\mu dH\Eqno$$

Varying $q$, $H$ and $\mu$ immediately gives
$$\sp f d\rho +\sp \gg  H\wd (i_a\lambda^a)=0\Eqno$$\label\qvar
$$ (i^a\lambda_a)\gg=d\mu\Eqno$$\label\Hvar
$$d H=0\Eqno$$\label\muvar
where $\rho=\rho^a{}_a$.
In order to evaluate \qvar\  and \Hvar\ further we must solve for the forms
$\rho$ and $i_a\lambda^a$.
This is achieved by analysing the equation obtained from the connection
$\L ab$ variations
$$ D*(e_a\wd e^a) +\lambda_a\wd e^b -\rho^b{}_a=0\Eqno$$\label\connvar
The first term
may be expressed in terms of the torsion and non-metricity forms:
$$D*(e_a\wd e^a)=D(\eta_{ac}*( e^c\wd e^a))=$$
$$=-2 Q_{ac}\wd * (e^c\wd e^b) +\eta_{ac}*(e^c\wd e^b\wd e^d)\wd T_d$$
and using \tor\ and \nonmet\ one finds
$$D*(e_a\wd e^a)=-2 df \wd *(e_a\wd e^b)+\gg  *(e_a\wd e^b\wd e^c)\wd i_cH$$
The symmetric and antisymmetric parts of \connvar\  now give
$$\l_a\wd e_b -\l_b\wd e_a=4 df\wd *(e_a\wd e_b) -2\gg  *(e_a\wd e_b\wd e_c)\wd
i^c H$$
$$\l_a\wd e_b +\l_b\wd e_a=2\rho_{ba}$$
Tracing the symmetric part yields
$$\rho=\l_a\wd e^a$$
while the anti-symmetric part yields after some calculation
$$i_a\l^a=3\gg *H$$
Hence
$$\rho=-12 * df\Eqno$$
and
$$\l_a=\gg  e_a\wd *H-4i_a*df\Eqno$$
Thus with the choice
$\gg =kq$, $f=\frac k2 ln(q)$ $k^2=\frac {\eta}{6} >0$
the equations \Hvar\  and \qvar\ become
 $$\frac {d*dq}{q^2} -\frac{dq \wd *dq}{q^3} -q H\wd * H=0\Eqno$$\label\qeqn
$$d*(q^2 H)=0\Eqno$$
In terms of the axion field,
$dH=0$ implies
 $$d(\frac{*d\ax}{q^2})=0\Eqno$$
and \qeqn\ becomes
$$d*dq-\frac{dq \wd *dq}{q}+\frac{d\ax\wd *d\ax}{q}=0\Eqno$$
These are the correct field equations for the axion and dilaton predicted
by low energy effective string actions.

Finally we examine the metric variations induced by varying the coframes.
With the aid of the identity
$$ \l^a\wd \delta_e(i_aH)=-\delta^a\wd i_b(\lambda^b\wd i_aH)$$
the Einstein equation follows from the action \action\ as
$$\R bc (\Lambda)\wd *(e_b\wd e^c\wd e_a) +D(\lambda)\lambda_a +kq
i_b(\lambda^b\wd i_aH)=0\Eqno$$\label\eineqn
It is a somewhat tedious calculation to decompose this into Levi-Civita
components.
The first term becomes
$$\R bc (\Lambda)\wd *(e_b\wd e^c\wd e_a) =
\R bc (\omega)\wd *(e_b\wd e^c\wd e_a)
+ke_a\wd d*(q H) -\frac {2k dq \wd i_a * dq}{q^2} +\frac{2k D(\omega)i_a
*dq}{q}$$
$$+\frac{k^2}{4}(-2q^2 i_aH \wd *H +6q^2 H\wd i_a *H +\frac{2}{q^2} dq \wd
i_a *dq -\frac{6}{q^2} i_adq\wd *dq)\Eqno$$
The second and third terms are respectively
$$D(\lambda)\lambda_a=-ke_a\wd d(q*H)+\frac{2k}{q^2} dq \wd i_a*dq
-\frac{2k}{q}D(\omega)(i_a*dq)$$
$$+k^2q^2 i_aH\wd *H +k^2 e_a\wd dq \wd *H+k^2 i^bi_a H\wd i_b*dq +\frac{3
k^2}{q^2}i_adq *dq +\frac{k^2}{q^2} dq\wd i_a *dq\Eqno$$
$$ kq i_b(\lambda^b\wd i_aH)= k^2q^2 i_aH \wd * H-2k^2 i^b *dq \wd i_b i_a
H\Eqno$$
Thus \eineqn\ may be written
$$ \R bc (\omega)\wd *
(e_b\wd  e^c\wd e_a)+\frac{\eta}{4}
q^2 \tau_a[H] +\frac{\eta}{4q^2}\tau_a[q]=0\Eqno$$\label\eineqnn
where the stress forms are
$$\tau_a[H]\equiv H\wd i_a *H +i_aH\wd *H$$
$$\tau_a[q]\equiv dq\wd i_a *dq +i_a dq\wd *dq$$
String theory couplings correspond to $\eta=1$.
This confirms that the field equations derived from the non-Riemannian
Einstein-Hilbert action above coincide with the axi-dilaton sector of the
low energy effective action derived from string theory.

\Section{Discussion}
We have expressed the axi-dilatonic sector of low energy string theory in
terms of a geometry with torsion and a metric gradient. This formulation
emphasises
the geometrical nature of the axion and dilaton fields and raises questions
about the most appropriate geometry for the discussion of physical
phenomena involving these fields.
In particular if the quanta of these fields can give rise to classical
particles then their gravitational interactions may correspond to
autoparallels of this non-Riemannian connection  rather than the geodesics
corresponding to the Levi-Civita connection.

\Section{ Acknowledgment}

The authors are grateful to EPSRC for providing facilities at the 15-th UK
Institute for Theoretical High Energy Physicists   at the University of
Southampton where this work was begun and
to the Human Capital and Mobility Programme of the European Union for partial
support. TD is grateful
to the School of Physics and Materials,
University of Lancaster, UK  for its hospitality.


\References

\bye